\documentclass[12pt]{article} \usepackage{amssymb,latexsym}
\usepackage{amsmath}
 
\columnwidth\textwidth

\begin{document}

\newtheorem{df}{Definition} \newtheorem{thm}{Theorem} \newtheorem{lem}{Lemma}
\newtheorem{rl}{Rule}
\begin{titlepage}
 
\noindent
 
\begin{center} {\LARGE On the theory of quantum measurement} \vspace{1cm}

P. H\'{a}j\'{\i}\v{c}ek\\ Institute for Theoretical Physics \\ University of
Bern \\ Sidlerstrasse 5, CH-3012 Bern, Switzerland \\ hajicek@itp.unibe.ch

\vspace{1cm}

June 2013 \\ \vspace{1cm}
 
PACS number: 03.65.Ta
 
\vspace*{2cm}
 
\nopagebreak[4]
 
\begin{abstract} The notion of state reduction employed by the standard
quantum theory of measurement is difficult to accept for two reasons: It
leaves open where and when the reduction takes place and it does not give any
objective conditions under which the reduction occurs. Some recently published
ideas on this problem are developed an improved. The disturbance of
measurement due to identical particles in the environment is shown to make any
POV measure non-measurable. Truncated POV (TPOV) measures are introduced that
can be measurable if object systems satisfy the additional requirement of
having separation status. The separation status is generalised from domain of
space to domain of phase space. Starting from the previously introduced
distinction between ancillas, screens and detectors, further study of
experiments suggests that a thermodynamic mixing within a detector or a screen
and the consequent loss of separation status is the objective condition for
the occurrence of the state reduction. The conjecture is simple, specific and
testable. The theory is illustrated by a model of a real measurement.
\end{abstract}

\end{center}

\end{titlepage}

\section{Introduction} It is well known that the quantum theory of measurement
is in an unsatisfactory state \cite{BLM,Ghirardi}. For example, the ideas of
quantum decoherence theory \cite{schloss} that have brought some progress does
not solve the problem of quantum measurement without any additional
assumptions such as Everett interpretation \cite{steveW,WM}.

A measurement on microscopic systems can be split into preparation and
registration. Registration devices are called {\em meters}. We observe that
the process of registration has the following strange and fascinating
properties:
\begin{enumerate}
\item Registered value $r$ is only created by the interaction of the object
system with the meter during the registration. Unlike the measurement in a
classical theory, registrations do not reveal already existing values. There
has been a lot of work on this feature since the beginnings of quantum
mechanics and it is very well confirmed by a number of theoretical and
experimental results (contextuality \cite{bell4,kochen}, Bell inequalities
\cite{bell1}, Hardy impossibilities \cite{hardy}, Greenberger-Horne-Zeilinger
equality \cite{GHZ}, etc.).
\item As a rule, repeated experiments give different values $r$ from a
well-defined set of possible alternatives ${\mathbf R}$. Each outcome is thus
created with probability ${\mathrm P}_r$ such that $\sum_{r\in{\mathbf
R}}{\mathrm P}_r = 1$. The resulting randomness, or the so-called QM {\em
indeterminism} occurs only during registrations. This is not in good harmony
with other quantum processes, which are described by Schr\"{o}dinger
equation in a deterministic way.
\item There are correlations between values registered by distant meters. As
the values are only created by registrations, a spooky action at a distance
between the meters turns out to be necessary. This is called QM {\em
non-locality} and it again seems to appear only via registrations (e.g.,
Einstein-Podolski-Rosen experiment \cite{EPR}, Bell has proved such
non-locality in the registration of a single particle \cite{bell2}).
\end{enumerate} These properties are not logically
self-contradictory. Moreover, they are testable and have been confirmed by
numerous experiments as well as theoretical analysis. For many physicists,
however, they are unacceptable for taste and traditional reasons. Our
standpoint is that they can be taken as ``facts of life''. However, they give
us an additional motivation to concentrate research on the phenomenon of
registration.

In several recent papers
\cite{HT,hajicek1,hajicek2,survey,hajicek3,hajicek4,models,hajicek5}, some new
ideas about quantum mechanics were proposed. The present paper is a
continuation of our work on the quantum measurement
\cite{hajicek2,hajicek4,models}. The main strategy has been a return to
physics: to observe carefully what happens in real experiments and to select
some general hypotheses, which have then an empirical rather than a
speculative character.

In \cite{hajicek2} it has been shown that quantum mechanical theory implies a
strong disturbance of the registration of any observable such as position,
momentum, energy, spin and orbital angular momentum on any quantum system
${\mathcal S}$. The origin of the disturbance is the existence of systems of
the same type as ${\mathcal S}$ in the world (not just in a neighbourhood of
${\mathcal S}$). For example, according to the standard quantum mechanics,
measurements of these observables on an electron is impossible because of the
existence of other electrons. This theoretical observation clearly contradicts
the long and successful praxis of experimenting. The fact that such
disturbances do not occur in real measurements must then be understood as a
proof that the current ideas on observables and registrations need some
corrections.

In \cite{hajicek2,hajicek4,models}, an attempt at a correction of measurement
theory is described that focuses on disturbances due to {\em remote}
particles. For example, the registration of a spin operator of an electron
prepared in our laboratory had to be (theoretically) disturbed by an electron
prepared in a distant laboratory. The idea was that it is not the spin
operator that is really measured, but a different observable that can be
constructed from the spin operator by some process of localisation. Such
constructs have been called $D$-local observables, $D$ being some region of
space. In order that $D$-local observables be measurable on, say, an electron,
the electron must be prepared in such a way that the influence of all other
electrons on the measurement apparatus inside a region $D$ of space is
negligible (e.g., the wave functions of all these electrons practically vanish
in $D$). We say then that the electron has a separation status $D$. On such a
basis, a whole general\footnote{For example, we speak here about wave
functions for the sake of simplicity, but wave functions are not sufficiently
general in two respects: they represent pure states and refer to a particular
frame, the $Q$-representation.} mathematical theory has been constructed
\cite{models}.

Another important fact is that a particle prepared with a certain local
separation status looses the status if it arrives in a region of space where
its wave function has a non-zero overlap with wave functions of other
particles of the same type. Then, the particle itself does not make sense as
an individual system in a prepared state because there are only
(anti-)symmetrised states of the whole system. Thus, changes of separation
status can accompany quantum processes and are recognised as deep changes in
physics of the studied systems. This is a physical idea that is not
sufficiently supported by standard quantum mechanical formalism. Moreover. a
change of separation status can be understood as an objective fact that can be
observed but is itself independent of any observer.

A careful study of many real experiments in \cite{hajicek2,hajicek4} has shown
that all meters contain macroscopic detectors and screens. Further observation
are that the registration processes include separation status losses in the
detectors and screens on one hand and state reductions on the other (screens
are studied in \cite{models}). This motivates us to state a general rule:
Changes of separation status are associated with state reductions. The form of
the state reduction is then determined by the structure of the
experiment. Thus, the state reduction occurs at a well-defined place and time
and the reduction frame is also determined.

That mean, of course, that Schr\"{o}dinger equation is not valid for
changes of separation status. This is rather surprising. It is difficult to
believe that the disturbance of measurement due to identical particle has
anything to do with the problem of quantum measurement because the standard
understanding of quantum theory does not lead to any idea in this
direction. However, the hypothesis is empirical: it is not derived by some
theoretical procedure and is only justified by observations. It is specific
and testable.

The resulting theory, which is described with many detail in \cite{models},
suggests the way in which the quantum measurement theory could and ought to be
corrected. However, it represents an idealised model whose practical
applications are limited. On the one hand, it only focuses on the space
aspects of quantum systems working exclusively with regions of the eigenspace
of the position operator and so violates the transformation symmetry of
quantum mechanics. On the other, the spatial separation status is rather
difficult to be prepared. We can, e.g., never achieve perfect vacuum in the
cavities where the experiment is done.

One can wonder whether the separation of particles in the momentum space could
play a similar role as that in the position one studied in \cite{models}. In
fact, it is straightforward to built up a formalism in the momentum space that
is completely analogous to the formalism in the position space. To see the
physical meaning of such a construction, imagine that the detector used in an
experiment has an energy threshold $E_0$. Then the particles with energy lower
than the threshold cannot influence the detector. It is easy to achieve
momentum separation status and, in fact, most measurements work exactly in
this way. It is the main purpose of the present paper to go a step further and
to give a suitable generalisation of separation status and of measurable
observables, using regions of a phase space rather than regions of the
position space.

The plan of the paper is as follows. Section 2 gives a brief account of the
standard theory of quantum measurement that is being used today for analysis
of real experiments. In this way, it will be possible to make explicit all
changes that the present paper contains. Section 3 introduces the notion of
truncated POV measures and justifies the proposal that these quantities must
be used for description of real experiments instead of POV measures of the
standard theory. This is a correction to our previous papers, which worked
with certain POV measures. Section 4 contains a generalisation of our previous
theory of separation status. Instead of defining the separation status as a
domain of the coordinate or momentum eigenspaces, it first defines an
approximate extent of a quantum system as a domain the Cartesian product of
the two eigenspaces. The notion of extent is then used to define a new kind of
separation status that seems to be more satisfactory than the old one. Section
5 recapitulates and reformulates some older ideas using also the new notion of
separation status. First, ancillas, screens and detectors of a given meter
must be distinguished by the analysis of the structure of the meter. Second,
the reading of a meter is postulated to be a signal from a detector. Third,
detected systems lose their separation status within screens and
detectors. Fourth, Schr\"{o}dinger equation does not hold for changes of
separation status and must be replaced by new rules. Section 6 describes a
simple model of the Stern-Gerlach measurement within our theory showing how
the new rules are to be understood. The last section gives a summary of the
paper.

\section{The standard theory of measurement} In this section, we give a short
review of the standard theory of measurement as it is employed in the analysis
of many measurements today and as it is described in, e.g.,
\cite{WM,bragin,svensson}. The emphasis is on being close to experiments and
on physical meaning rather than on mathematical formalism.

The standard theory splits a measurement process into three steps.
\begin{enumerate}
\item Initially, the {\em object system} ${\mathcal S}$ on which the
measurement is to be done is prepared in state ${\mathsf T}_{{\mathcal
S}}^{\text{in}}$ and the {\em meter} ${\mathcal M}$, that is the aparatus
performing the measurement, in state ${\mathsf T}_{{\mathcal
M}}^{\text{in}}$. These two preparations are independent so that the composite
${\mathcal S} + {\mathcal M}$ is then in state ${\mathsf T}_{{\mathcal
S}}^{\text{in}} \otimes {\mathsf T}_{{\mathcal M}}^{\text{in}}$. ${\mathsf
T}_{{\mathcal S}}^{\text{in}}$ and ${\mathsf T}_{{\mathcal M}}^{\text{in}}$
are {\em state operators} (sometimes also called density matrices).
\item An interaction between ${\mathcal S}$ and ${\mathcal M}$ suitably
entangles them. This can be theoretically represented by unitary map ${\mathsf
U}$, called {\em measurement coupling}, that describes the evolution of system
${\mathcal S} + {\mathcal M}$ during a finite time interval. Hence, at the end
of the time interval, the composed system is supposed to be in state
$$
{\mathsf U} ({\mathsf T}_{{\mathcal S}}^{\text{in}} \otimes {\mathsf
T}_{{\mathcal M}}^{\text{in}}) {\mathsf U}^\dagger\ .
$$
\item Finally, {\em reading} the meter gives some {\em definite} value $r$ of
the measured quantity. If the same measurements are repeated more times
independently from each other, then all readings form a set, $r \in {\mathbf
R}$. ${\mathbf R}$ is not necessarily the spectrum of an observable (s.a.\
operator), in particular, it need not contain only real numbers (${\mathbf R}$
need not be a subset of ${\mathbb R}$). The experience with such repeated
measurements is that each reading $r \in {\mathbf R}$ occurs with a definite
probability, ${\mathrm P}_r$.
\end{enumerate} One of the most important assumptions of the standard theory
is that, after the reading of the value $r$, the object system ${\mathcal S}$
is in a well-defined state,
$$
{\mathsf T}_{{\mathcal S}r}^{\text{out}}\ ,
$$
called {\em conditional} or {\em selective} state. This is a generalisation of
Dirac postulate:
\begin{quote} A measurement always causes a system to jump in an eigenstate of
the observed quantity.
\end{quote} Such a measurement is called {\em projective} and it is the
particular case when ${\mathsf T}_{{\mathcal S}r}^{\text{out}} =
|r\rangle\langle r|$ where $|r\rangle$ is the eigenvector of a s.a.\ operator
for a non-generated eigenvalue $r$.

The average of all conditional states after registrations, a {\em proper
mixture},
$$
\sum_r {\mathrm P}_r {\mathsf T}_{{\mathcal S}r}^{\text{out}}\ ,
$$
is called {\em unconditional} or {\em non-selective} state. It is described as
follows: ``make measurements but ignore the results''. One also assumes that
$$
\sum_r {\mathrm P}_r {\mathsf T}_{{\mathcal S}r}^{\text{out}} = tr_{\mathcal
M}\Big({\mathsf U} ({\mathsf T}_{{\mathcal S}}^{\text{in}} \otimes {\mathsf
T}_{{\mathcal M}}^{\text{in}}) {\mathsf U}^\dagger\Big)\ ,
$$
where $tr_{\mathcal M}$ denotes a partial trace defined by any orthonormal
frame in the Hilbert space of the meter.

In the standard theory, the reading is a mysterious procedure. If the meter is
considered as quantum system then to observe it, another meter seems to be
needed, to observe this, still another is needed and the resulting series of
measurements is called {\em von-Neumann chain}. At some (unknown) stage
including the processes in the mind of observer, there is the so-called {\em
Heisenberg cut} that gives the definite value $r$. Moreover, the conditional
state cannot, in general, result by a unitary evolution. The transition
$$
tr_{\mathcal M}\Big({\mathsf U} ({\mathsf T}_{{\mathcal S}}^{\text{in}}
\otimes {\mathsf T}_{{\mathcal M}}^{\text{in}}) {\mathsf U}^\dagger\Big)
\mapsto {\mathsf T}_{{\mathcal S}r}^{\text{out}}
$$
in each individual registration is called ``the first kind of dynamics''
\cite{JvN} or ``state reduction'' or ``collapse of the wave function''. We
will use the name ``state reduction''.

The idea of state reduction is difficult to accept for two reasons. First, the
time and location of the Heisenberg cut is not known. Thus, the theory is
incomplete. Second, if there are two different kinds of dynamics, there ought
to be also objective conditions under which each of them is applicable. At the
present time, no such objective conditions are known. For example, for the
state reduction, the condition of the presence of an observer is not objective
and the condition that a quantum system interacts with a macroscopic system is
not necessary.

The standard theory describes a general measurement mathematically by two
quantities. The first is a {\em state transformer} ${\mathcal
O}_r$. ${\mathcal O}_r$ enables us to calculate ${\mathsf T}_{{\mathcal
S}r}^{\text{out}}$ from ${\mathsf T}_{{\mathcal S}}^{\text{in}}$ by
$$
{\mathsf T}_{{\mathcal S}r}^{\text{out}} = \frac{{\mathcal O}_r({\mathsf
T}_{{\mathcal S}}^{\text{in}})}{tr\Big({\mathcal O}_r({\mathsf T}_{{\mathcal
S}}^{\text{in}})\Big)}\ .
$$
${\mathcal O}_r$ is a so-called {\em completely positive map} that has the
form \cite{kraus}
\begin{equation}\label{kraus} {\mathcal O}_r({\mathsf T}) = \sum_k {\mathsf
O}_{rk}{\mathsf T}{\mathsf O}_{rk}^\dagger
\end{equation} for any state operator ${\mathsf T}$, where ${\mathsf O}_{rk}$
are some operators satisfying
$$
\sum_{rk} {\mathsf O}_{rk}^\dagger{\mathsf O}_{rk} = {\mathsf 1}\ .
$$
Equation (\ref{kraus}) is called {\em Kraus representation}. A given state
transformer ${\mathcal O}_r$ does not determine, via Eq. (\ref{kraus}), the
operators ${\mathsf O}_{rk}$ uniquely.

The second quantity is a {\em probability operator} ${\mathsf E}_r$ (often
called ``effect'') giving the probability to read value $r$ by
$$
{\mathrm P}_r = tr\Big({\mathcal O}_r({\mathsf T}_{{\mathcal
S}}^{\text{in}})\Big) = tr({\mathsf E}_r {\mathsf T}_{{\mathcal
S}}^{\text{in}})\ .
$$
The set $\{{\mathbf E}_r\}$ of probability operators ${\mathsf E}_r$ for all
$r \in {\mathbf R}$ is called {\em probability operator valued (POV) measure}
(often called ``positive operator valued''). Every POV measure satisfies two
conditions: positivity,
$$
{\mathsf E}_r \geq {\mathsf 0}\ ,
$$
for all $r \in {\mathbf R}$, and normalisation,
$$
\sum_{r\in{\mathbf R}} {\mathsf E}_r = {\mathsf 1}\ .
$$

One can show that ${\mathcal O}_r$ determines the probability operator
${\mathsf E}_r$ by
$$
{\mathsf E}_r = \sum_{k} {\mathsf O}_{rk}^\dagger{\mathsf O}_{rk}\ .
$$

The definition of POV measures that is usually given is more general:
${\mathsf E}(X)$ is a function on the Borel subsets $X \subset {\mathbf
R}$. The formalism that we introduce in the present paper can be easily
generalised in this way.

In the standard theory, the state transformer of a given registration contains
all information that is necessary for further analysis and for classification
of measurements. Such a classification is given in \cite{WM}, p.\ 35. Thus,
the formalism of the state transformers and POV measures can considered as the
core of the standard theory.

The standard quantum mechanics defines observables of a system ${\mathcal S}$
as the self-adjoint operators on the Hilbert space of ${\mathcal S}$. Some
mathematical physicists (e.g., Ludwig, Bush, Lahti and Mittelstaed) define
observables as POV measures. The spectral measures of s.a.\ operators are POV
measures and in this sense, the definition is a generalisation of the standard
definition.

The authors of \cite{WM} return to the standard nomenclature and distinguish
observables from POV measures. Observables are used in many ways, in
particular to construct POV measures, but they are only indirectly related to
measurements. In fact, only a special class of measurement can then be called
``measurement of an observable''. This is the case when all probability
operators ${\mathsf E}_r$ of a POV measure are functions of an observable
(\cite{WM}, p.\ 38). There are important measurements that do not satisfy this
condition. This standpoint is not generally accepted and is not shared by our
previous papers (see, e.g., \cite{models}), but it seems very reasonable and
we shall adopt it here. It will turn out that even POV measures can be related
to real measurements only indirectly because of the disturbance of measurement
due to identical particles.

\section{Truncated POV measures} Let us first briefly recall the argument of
\cite{hajicek2} about the disturbance of registration by identical
particles. Consider two distant laboratories, $A$ and $B$, and suppose that
each of them prepares an electron in states $\psi(\vec{x}_A)$ and
$\phi(\vec{x}_B)$, respectively (we are leaving out the spin indices and we
work in $Q$-representation for the sake of simplicity). Then, the everyday
experience shows that $A$ can do all manipulations and measurements on its
electron without finding any contradictions to the assumption that the state
is $\psi(\vec{x}_A)$. Analogous statements hold about $B$.

However, according to the standard quantum theory, the state of the two
particles must be
\begin{equation}\label{symstate} 2^{-1/2}\bigl(\psi(\vec{x}_A)\phi(\vec{x}_B)
- \phi(\vec{x}_A)\psi(\vec{x}_B)\bigr)\ .
\end{equation} Suppose next that $A$ makes a measurement of the position of
the electron. Standard quantum mechanics associates position observable with
the multiplication operator $\vec{x}_A$ for $A$ electron and with a
symmetrised multiplication operator
\begin{equation}\label{symobs} \vec{x}_A + \vec{x}_B
\end{equation} for the two electrons because the meter cannot distinguish the
contributions of two identical particles from each other. Hence, the average
of position measurement must be
\begin{equation}\label{aver} \int_{\mathbb
R}d^3x_A\,\vec{x}_A|\psi(\vec{x}_A)|^2 + \int_{\mathbb
R}d^3x_B\,\vec{x}_B|\phi(\vec{x}_B)|^2
\end{equation} which differs from what one would expect if the state of the
electron were just $\psi(\vec{x}_A)$, and the difference even increases with
the distance of the laboratories.

A natural way out of this contradiction between the standard quantum mechanics
and experience has been suggested by Peres \cite{peres}. The meters that can
be used by laboratory $A$ clearly cannot react to particles with wave
functions that practically vanish within the laboratory, which is true for
$\phi(\vec{x})$, at least approximately and for some time. Then any observable
${\mathsf O}$ measured by such a meter satisfies $\langle \phi|{\mathsf
O}|\phi\rangle = 0$ and the corresponding second term in the equation
analogous to (\ref{aver}) vanishes. However, the unexpected consequence of
this explanation is that the position observable registered by the device
cannot be $\vec{x}_A$!

Moreover, the device cannot measure any of the standard observables such as
energy, momentum, spin, etc. A general proof that such a device cannot measure
any POV measures goes as follows. Suppose that ${\mathsf E}_r$ is a POV
measure of quantum system ${\mathcal S}$. In order that state ${\mathsf T}$ of
any system indistinguishable from ${\mathcal S}$ in the environment does not
disturb the registration of ${\mathsf E}_r$, the probability that the
measurement of ${\mathsf E}_r$ on ${\mathsf T}$ gives any result $r$ must be
zero. For that, the following condition is sufficient and necessary:
\begin{equation}\label{POVnondist} tr({\mathsf T}{\mathsf E}_r) = 0 \quad
\forall r\ .
\end{equation} However, the normalisation condition implies
$$
\sum_r tr({\mathsf T}{\mathsf E}_r) = 1\ ,
$$
which contradicts (\ref{POVnondist}). The genuine meters must be such that
they do not react to some states.

If not POV measures, which quantities describe registrations? Let us define
{\em truncated POV measures} (TPOV measures) as follows. In general, any given
experiment Exp on system ${\mathcal S}$ using meter ${\mathcal M}$ works with
a limited set ${\mathbf T}_{\text{Exp}} = \{{\mathsf T}_1,{\mathsf
T}_2,\ldots,{\mathsf T}_K,\}$ of states in which ${\mathcal S}$ is prepared
before registrations. We assume that there is subspace ${\mathbf
H}_{\text{Exp}}$ of Hilbert space ${\mathbf H}$ of ${\mathcal S}$ satisfying
two conditions. First,
\begin{equation}\label{hexp} {\mathsf \Pi}[{\mathbf H}_{\text{Exp}}]{\mathsf
T}{\mathsf \Pi}[{\mathbf H}_{\text{Exp}}] = {\mathsf T}
\end{equation} for all ${\mathsf T} \in {\mathbf T}_{\text{Exp}}$, where
${\mathsf \Pi}[{\mathbf H}_{\text{Exp}}] : {\mathbf H} \mapsto {\mathbf
H}_{\text{Exp}}$ is an orthogonal projection. Second, ${\mathbf
H}_{\text{Exp}}$ is minimal, that is any subspace of ${\mathbf H}$ that
satisfies (\ref{hexp}) must contain ${\mathbf H}_{\text{Exp}}$. In fact, for
most experiments, ${\mathbf H}_{\text{Exp}}$ is a finite-dimensional subspace
of ${\mathbf H}$.
\begin{df} Any TPOV measure associated with experiment Exp is a set
$\{{\mathsf E}'_r\}$of s.a.\ operators satisfying
$$
{\mathsf E}'_r \geq {\mathsf 0}
$$
for all $r \in {\mathbf R}$ and
$$
\sum_r {\mathsf E}'_r = {\mathsf \Pi}[{\mathbf H}_{\text{Exp}}]\ .
$$
\end{df} {\bf Example} Let ${\mathsf E}_r$ be POV measure. Then:
$$
{\mathsf E}'_r = {\mathsf \Pi}[{\mathbf H}_{\text{Exp}}]{\mathsf E}_r{\mathsf
\Pi}[{\mathbf H}_{\text{Exp}}]
$$
is a {\em TPOV measure}.

We have the desired property: states ${\mathsf T}$ annihilated by ${\mathsf
\Pi}[{\mathbf H}_{\text{Exp}}]$ satisfy $tr({\mathsf T}{\mathsf E}'_r) = 0$
for any $r$. An example of a TPOV measure is described in Sec.\ 6.

\section{Separation status} The foregoing section introduced quantities that
need not be disturbed by identical particles during registrations. However,
more conditions must be satisfied in order that a registration be not
disturbed. Let us introduce further mathematics. First, we need some measure
of the {\em extent} of a quantum system.

\begin{df}\label{dfextent} Let ${\mathcal S}_\tau$ be a system of $N$
particles of type $\tau$ in state ${\mathsf T}(t)$ at time $t$. Let ${\mathsf
a}_k$ be an observable of the $k$-th particle. Let
\begin{equation}\label{kubar} \bar{a} = tr\left({\mathsf
T}\frac{\sum_k{\mathsf a}_k}{N}\right)
\end{equation} and
\begin{equation}\label{deltaku} \Delta a = \sqrt{tr\left({\mathsf
T}\frac{\sum_k({\mathsf a}_k - \bar{a})^2}{N}\right)}\ .
\end{equation} The {\em extent} $\text{Ext}({\mathsf T})$ of ${\mathsf T}$ is
the domain of ${\mathbb R}^6$ defined by the Cartesian product,
\begin{equation}\label{extent} \text{Ext}({\mathsf T}) = \prod_i^3
\Big(\bar{x}^i - \Delta{x}^i,\bar{x}^i + \Delta{x^i}\Big) \prod_j^3
\Big(\bar{p}^j - \Delta{p^j},\bar{p}^j + \Delta{p^j}\Big)\ ,
\end{equation} where $\bar{x}^i$ and $\Delta{x^i}$ are determined by Equations
(\ref{kubar}) and (\ref{deltaku}) for ${\mathsf a}_k = {\mathsf x}_k^i$,
${\mathsf x}_k^i$ being the $i$-component of the position operator of $k$'s
particle in ${\mathcal S}_\tau$ and similarly for $\bar{p}^j$ and
$\Delta{p^j}$, ${\mathsf a}_k = {\mathsf p}_k^j$, ${\mathsf p}_k^j$ being the
$j$-component of the momentum operator of $k$'s particle in ${\mathcal
S}_\tau$.
\end{df} For example, consider two bosons in state ${\mathsf T} =
|\Psi\rangle\langle\Psi|$, where
$$
|\Psi\rangle = \frac{1}{\sqrt{2}}(|\psi 1\rangle \otimes |\phi 2\rangle +
|\phi 1\rangle \otimes |\psi 2\rangle)\ ,
$$
$|\psi\rangle$ and $|\phi\rangle$ are two vector states in the common Hilbert
space of the two bosons satisfying $\langle\psi|\phi\rangle = 0$ and the
symbol $|\psi k\rangle$ means that the state $|\psi\rangle$ is occupied by the
$k$-th particle. A short calculation gives
$$
\bar{a} = \frac{\langle\psi|{\mathsf a}|\psi\rangle + \langle\phi|{\mathsf
a}|\phi\rangle}{2}
$$
and
$$
\Delta a = \sqrt{\frac{1}{2}\left(\Delta^2_{\psi} a + \Delta^2_{\phi} a +
\frac{1}{2}(\langle\psi|{\mathsf a}|\psi\rangle - \langle\phi|{\mathsf
a}|\phi\rangle)^2\right)}\ ,
$$
where
$$
\Delta^2_{\psi} a = \langle\psi|{\mathsf a}^2|\psi\rangle -
\langle\psi|{\mathsf a}|\psi\rangle^2
$$
and
$$
\Delta^2_{\phi} a = \langle\phi|{\mathsf a}^2|\phi\rangle -
\langle\phi|{\mathsf a}|\phi\rangle^2\ .
$$
We can see that the extent includes not only the ``sizes'' ($\Delta_{\psi} a$)
of individual particles but also the ``distances'' ($|\langle\psi|{\mathsf
a}|\psi\rangle - \langle\phi|{\mathsf a}|\phi\rangle|$) of different particles
in ${\mathcal S}_\tau$.

\begin{df}\label{sepst} Given a system ${\mathcal S}$, let ${\mathcal S}_\tau$
be the subsystem of ${\mathcal S}$ containing all particles in ${\mathcal S}$
of type $\tau$. Similarly, let ${\mathcal E}_\tau$ be the subsystem of
environment of ${\mathcal S}$ that contains all particles of type $\tau$. We
say that ${\mathcal S}$ has a {\em separation status} if the extents of
${\mathcal S}_\tau$ and ${\mathcal E}_\tau$ have empty intersection for all
$\tau$.
\end{df}

To give some physical interpretation to this formalism, consider meter
$\mathcal M$ that is able to register systems of the same type as ${\mathcal
S}$. Then, in order to be registered by $\mathcal M$, ${\mathcal S}$ has to be
at some time inside $\mathcal M$ and its kinetic energy must lie in the
interval $(E_0,\infty)$ defined by threshold $E_0$ of the meter. The direction
of the momentum must lie in the interval in which ${\mathcal S}$ must arrive
at the meter in order to be registered. These are condition on the extent of
${\mathcal S}_\tau$ for all types $\tau$.

We assume first: every measurement on a system ${\mathcal S}$ with no
separation status will be disturbed by particles in its neighbourhood. Second,
experiments on ${\mathcal S}$ can be arranged so that they will be only
negligibly disturbed by environment particles if ${\mathcal S}$ has a
separation status. The TPOV of the registration will then practically not
react to states with very different extent.

Separation status has been defined in \cite{hajicek2} as a region of
space. The space ${\mathbb R}^6$ used in the definitions above can be
considered as the phase space of one classical particle, and this is why we
can say that the present section generalises the old definition from regions
of space to regions of phase space. However, one ought to keep in mind that
${\mathbb R}^6$ is the phase space of the considered system only in very few
cases.

We can interpret what has been said as yet as follows. Standard quantum
mechanics as it is usually presented seems incomplete:
\begin{enumerate}
\item It admits only two separation statuses for any system ${\mathcal S}$:
\begin{enumerate}
\item ${\mathcal S}$ is isolated. Then all states of ${\mathcal S}$ have
separation status and all s.a.\ operators are measurable.
\item ${\mathcal S}$ is a member of a larger system containing particles
identical to ${\mathcal S}$. Then there are no individual physical states and
observables for ${\mathcal S}$.
\end{enumerate}
\item It ignores the existence of separation-status changes.
\end{enumerate} However, separation-status changes have two important
features:
\begin{enumerate}
\item They are objective phenomena that happen independently of any observer,
and can be distinguished from other quantum mechanical processes.
\item Losses of separation status seem to be associated with state
reductions. This gives us some hope that state reductions would indeed occur
only if some objective conditions were satisfied.
\end{enumerate}

\section{Theory of meter reading} Let us now show in more detail how
registered systems lose their separation status in meters.

In many modern experiments, in particular in non-demolition and weak
measurements, but not only in these, the following idea is employed. The
object system $\mathcal S$ interact first with a microscopic system $\mathcal
A$ that is prepared in a suitable state. After $\mathcal S$ and $\mathcal A$
become entangled, $\mathcal A$ is subject to further registration and, in this
way, some information on $\mathcal S$ is obtained. No subsequent measurements
on $\mathcal S$ has to be made. The state of $\mathcal S$ is influenced by the
registration just because of its entanglement with $\mathcal A$. The auxiliary
system $\mathcal A$ is usually called {\em ancilla}.

It seems, however, that any registration on microscopic systems has to use
{\em detectors} in order to make features of microscopic systems visible to
humans. Detector is a macroscopic system containing {\em active volume}
$\mathcal D$ and {\em signal collector} $\mathcal C$ in thermodynamic state of
metastable equilibrium. Notice that the active volume is a physical system,
not just a volume of space. Interaction of the detected systems with $\mathcal
D$ triggers a relaxation process leading to macroscopic changes in the
detector that are called {\em detector signals}. For the theory of detectors,
see, e.g., \cite{leo,stefan}.

Study of various experiments suggests that one can distinguish between
ancillas and detectors within meters and that this distinction provides a
basis for the analysis of meters. To be suitable for this aim, we have to
modify a little the current notions of detector and ancilla. On the one hand,
detectors as defined above are more specific than what may be sometimes
understood as detectors. On the other, ancillas as defined above are more
general.

For example, consider a ionisation gas chamber that detects a particle
$\mathcal S$ so that $\mathcal S$ first enters the active volume $\mathcal D$
of the chamber and then $\mathcal S$ can leave $\mathcal D$ again and be
subject to further measurements. $\mathcal S$ interacts with several gas atoms
in the chamber that become ionised. This microscopic subsystem of several
atoms within the active volume can also be viewed as an ancilla $\mathcal
A$. $\mathcal A$ ``is detected'' subsequently by the rest of the detector,
that is, $\mathcal A$ interacts with $\mathcal D$ and $\mathcal C$ and is
involved in a process of relaxation that leads to a macroscopic electronic
signal.

Study of experiments suggests further that the measurements on ancillas needs
detectors. Thus we are lead to the following hypothesis
\cite{hajicek2}:\par\vspace*{.5cm}\noindent {\bf Pointer Hypothesis} {\itshape
Any meter for microsystems must contain at least one detector and every
reading of the meter can be identified with a signal from a
detector.} \par\vspace*{.5cm}\noindent This is a very specific assumption that
is, on the one hand, testable and, on the other, makes the reading of meters
less mysterious.

In the above example of ionisation chamber, the state of the ancilla that is
prepared by the interaction with the object system has, initially, a
separation status: it can be distinguished from other systems of atoms within
$\mathcal D$ and, therefore, registered without external disturbance. However,
in the process of interaction with $\mathcal D$ and $\mathcal C$ and the
relaxation process, its energy is dissipated and its position is smeared so
that it loses its separation status. We assume next:
\par\vspace*{.5cm}\noindent {\bf Active-Volume Hypothesis} {\itshape Active
volume ${\mathcal D}$ of the detector detecting system ${\mathcal S}'$
contains many particles in common with ${\mathcal S}'$. The state of
${\mathcal S}' + {\mathcal D}$ then dissipates so that ${\mathcal S}'$ loses
its separation status.} \par\vspace*{.5cm}\noindent Thermodynamic relaxation
is necessary to accomplish the loss. ${\mathcal S}'$ might be the objects
system or an ancilla of the original experiment.

Study of a number of real experiments \cite{hajicek4,models} suggests the
following: \par\vspace*{.5cm}\noindent {\bf Separation Status Hypothesis}
{\itshape Let the Schr\"{o}dinger equation for the composite ${\mathcal
S} + {\mathcal M}$ leads to a linear superposition of alternative evolutions
such that some of the alternatives contain loss of separation statuses of the
object system or ancilla(s). Then, there is a state reduction of the linear
superposition to the proper mixture of the alternatives.}
\par\vspace*{.5cm}\noindent The rule is general (it includes also separation
status loss in screens, see \cite{models}) and thus necessarily somewhat
vague. This is, however, analogous to any other general dynamical law: even
Schr\"{o}dinger equation is defined only in rough features and must be
set up for each case separately. It turns out that the state reduction is
uniquely determined by the structure of the experiment and the losses of
separation status. The Separation Status Hypothesis is again a testable
hypothesis.

Suppose that a microscopic system ${\mathcal S}$ is detected by a detector
${\mathcal D} + {\mathcal C}$ and that ${\mathcal S}$ loses its separation
status within ${\mathcal D} + {\mathcal C}$. Then, ${\mathcal S}$ ceases to be
an individual quantum system with its own physical states and observables. At
most, one can consider the subsystem ${\mathcal S}_+$ of ${\mathcal S} +
{\mathcal D} + {\mathcal C}$ that contains all particles of the same type as
those that were inside ${\mathcal S}$ originally. ${\mathcal S}_+$ contains
more particles than ${\mathcal S}$ does and is, as a rule, a macroscopic
system.

To explain the Separation Status Hypothesis, let us assume for the sake of
simplicity, we that ${\mathcal S}_+$ is a closed quantum system so that its
states and their standard evolution defined by Schr\"{o}dinger equation
(``formal evolution'' \cite{hajicek4,models}) make sense. This assumption is,
strictly speaking, incorrect because ${\mathcal S}_+$ is a macroscopic system
that cannot be isolated. However, the environment can be considered as
included, such an inclusion does not lead to really new phenomena and so such
an assumption does not necessarily lead to false conclusions.

The Hypothesis considers the standard evolution of ${\mathcal S}_+$, or a
larger system. After finding all cases of separation status losses, it
determines corrections to the standard evolution. This corrections are
suggested by real experiments and cannot be derived form standard quantum
mechanics.

In any case, the basic assumption of the old theory of quantum measurement
that the object system has a well-defined state after the registration is not
generally valid and the notion of state transformer does not make sense in
many important cases.

Observe that our proposals give the preparation and registration procedures
new importance with respect to, say, Copenhagen interpretation: they must
include changes of separation status.

\section{Stern-Gerlach story retold} In this section, we shall modify the
textbook description (e.g., \cite{peres}, p.\ 14) of the Stern-Gerlach
experiment utilising the above ideas.

A silver atom consists of 47 protons and 61 neutrons in the nucleus and of 47
electrons around it, but we consider only its mass-centre and spin degrees of
freedom and denote the system with these degrees of freedom by ${\mathcal
S}$. Let $\vec{\mathsf x}$ be its position, $\vec{\mathsf p}$ its momentum and
${\mathsf S}_z$ the $z$-component of its spin with eigenvectors $|j\rangle$
and eigenvalues $j \hbar/2$, where $j = \pm$.

Let ${\mathcal M}$ be a Stern-Gerlach apparatus with an inhomogeneous magnetic
field in a region $D$ that splits different $z$-components of spin of a silver
atom arriving in $D$ with a momentum in a suitable direction. Let a
scintillation-emulsion film with energy threshold $E_0$ be placed orthogonally
to the split beam. The scintillation emulsion is the active volume ${\mathcal
D}$ of ${\mathcal M}$ and it may be also the signal collector if the
scintillation events can be made directly visible.

First, let ${\mathcal S}$ be prepared at time $t_1$ in a definite
spin-component state
\begin{equation}\label{sin} |\vec{p},\Delta \vec{p}\rangle \otimes |j \rangle\
,
\end{equation} where $|\vec{p},\Delta \vec{p}\rangle$ is a Gaussian wave
packet so that ${\mathcal S}$ can be registered by ${\mathcal M}$ within some
time interval $(t_1,t_2)$. Let state (\ref{sin}) has a separation status at
$t_1$.

${\mathcal M}$ is in initial metastable state ${\mathsf T}_{\mathcal M}(t_1)$
at $t_1$.

Interaction of ${\mathcal S}$ with ${\mathcal M}$ is described by measurement
coupling ${\mathsf U}$. The time evolution within $(t_1,t_2)$ is:
$$
{\mathsf U}N{\mathsf \Pi}(|\vec{p},\Delta \vec{p}\rangle\langle \vec{p},\Delta
\vec{p}| \otimes |j \rangle \langle j |\otimes {\mathsf T}_{\mathcal
M}(t_1)){\mathsf \Pi}{\mathsf U}^\dagger = {\mathsf T}_{j}(t_2)\ ,
$$
where ${\mathsf \Pi}$ is antisymmetrisation on the Hilbert space of silver
atom part of ${\mathcal S} + {\mathcal M}$ and $N$ is a normalisation factor
because ${\mathsf \Pi}$ does not preserve normalisation. States ${\mathsf
T}_{j}(t_2)$ are determined by these conditions

This evolution includes a thermodynamic relaxation of ${\mathcal M}$ with
${\mathcal S}$ inside ${\mathcal D}$. States ${\mathsf T}_{j}(t_2)$ describe
subsystem ${\mathcal S}$ that has lost its separation status. Then, individual
states of ${\mathcal S}$ do not make sense: {\em neither the conditional state
nor the state transformer exist for ${\mathcal S}$}. (These notions are, in
fact, applicable only for some parts of some measurements with ancilla.)

State ${\mathsf T}_{j}(t_2)$ also describes detector signals. The signals will
be concentrated within one of two strips on the film, each strip corresponding
to one value of $j$.

Suppose next that the initial state of ${\mathcal S}$ at $t_1$ is
$$
|\vec{p},\Delta \vec{p}\rangle\left(\sum_j c_j |j\rangle\right)
$$
with
$$
\sum_j|c_j|^2 = 1\ .
$$

As it is linear, unitary evolution ${\mathsf U}$ gives
$$
N{\mathsf U}{\mathsf \Pi}\left[|\vec{p},\Delta \vec{p}\rangle
\langle\vec{p},\Delta \vec{p}| \otimes \left(\sum_jc_j|j \rangle\right)
\left(\sum_{j'}c_{j'}\langle j'|\right)\otimes {\mathsf T}_{\mathcal
M}(t_1)\right]{\mathsf \Pi}{\mathsf U}^\dagger = \sum_{jj'}c_jc^*_{j'}
{\mathsf T}_{jj'}(t_2)\ ,
$$
a quadratic form in $\{c_j\} \in {\mathbb C}^2$. Coefficients ${\mathsf
T}_{jj'}(t_2)$ of the form are operators on the Hilbert space of ${\mathcal S}
+ {\mathcal M}$.

The operator coefficients are state operators only for $j' = j$. From the
linearity of ${\mathsf U}$, it follows that
$$
{\mathsf T}_{jj}(t_2) = {\mathsf T}_{j}(t_2)\ .
$$

Now we postulate the following correction to the Schr\"{o}dinger equation
\begin{enumerate}
\item The loss of separation status of ${\mathcal S}$ disturbs the standard
quantum evolution so that, instead of
$$
\sum_{j,j'} c_jc^*_{j'} {\mathsf T}_{jj'}(t_2)\ ,
$$
state
$$
\sum_j |c_j|^2 {\mathsf T}_{j}(t_2)
$$
results.
\item States ${\mathsf T}_{j}(t_2)$ are uniquely determined by the
experimental arrangement: the measurement coupling and the losses of
separation status in the meter.
\item The sum is not only a convex combination but also a proper mixture of
the signal states ${\mathsf T}_{j}(t_2)$. That is, the system ${\mathcal S} +
{\mathcal M}$ is always in one particular state ${\mathsf T}_{j}(t_2)$ after
each individual registration and the probability for that is $|c_j|^2$.
\end{enumerate}

The described example is simple because the silver atoms are both the object
systems and components of the detector. If the detector contained no silver,
we would have to insert an intermediate step suggested by the fourth paragraph
of Section 5.

Stern-Gerlach experiment measures values of a truncated POV measure that
consist of two probability operators,
$$
{\mathsf E}_j = |\vec{p},\Delta \vec{p}\rangle\langle\vec{p},\Delta
\vec{p}|\otimes |j\rangle\langle j|\ ,
$$
where $j = \pm$. Clearly, the set $\{{\mathsf E}_j\}$ lives on a
two-dimensional subspace ${\mathbf H}_{\text{Exp}}$ of the Hilbert space of
the system ${\mathcal S}$ that is defined by the projection
$$
{\mathsf \Pi}[{\mathbf H}_{\text{Exp}}] = |\vec{p},\Delta
\vec{p}\rangle\langle\vec{p},\Delta \vec{p}|\ .
$$

\section{Conclusion} We have shown that the disturbance due to identical
particles makes the registration of any POV measures impossible. In articular,
none of the s.a.\ operators such as position, momentum, spin, angular momentum
or energy are measurable.

The explanation of why real measurements do not seem to be disturbed is,
first, that different quantities than POV measures are registered. As such
quantities, we have proposed TPOV measures. Second, the preparations of the
object systems satisfy an additional condition that is usually not
mentioned. To describe the condition, the notion of separation status has been
introduced in \cite{hajicek2}. Here, we have generalised the notion so that
some problems with the original notion disappear.

The next crucial observation is that the roles of ancilla and detector in
registrations must be distinguished from each other. We have then conjectured
that every meter contains at least one detector and meter readings are always
signals of detectors. Moreover, separation statuses are lost in detectors.

Finally, study of different kinds of real experiments show that the changes of
separation status are associated with state reduction. We assume that this is
a general empirical fact. In this way, the surprising connection between the
quantum theory of identical particles and the problem of quantum measurement
has been established.

What is called ``collapse of wave function'' can then be explained as state
degradation due to loss of separation status of the object system or an
ancilla by a thermodynamic relaxation process in a detector. Hence, the
collapse occurs under specific objective conditions and has a definite place
and time.

The correction to Schr\"{o}dinger equation is uniquely determined in each
case by the measurement coupling and the separation status losses. There is no
problem of preferred frame \cite{schloss}.

All conjectures made in this paper are testable.

\subsection*{Acknowledgements} The author is indebted to J\"{u}rg
Fr\"{o}hlich and Ji\v{r}\'{\i} Tolar for useful discussions.

\end{document}